\documentclass[12pt]{article}
\usepackage{amsmath, amssymb, slashed, epsf, color, graphicx}

\newcommand{\vev}[1]{ \left\langle {#1} \right\rangle }

\begin{document}

\begin{titlepage}
\begin{center}

\hfill IPMU--13--0078 \\
\hfill UT--13--14 \\
\hfill April 5, 2013

\vspace{1.5cm}
{\large\bf
Recent Result of the AMS-02 Experiment and \\
Decaying Gravitino Dark Matter in Gauge Mediation \\}

\vspace{2.0cm}
{\bf Masahiro Ibe}$^{(a, b)}$,
{\bf Sho Iwamoto}$^{(b)}$,
{\bf Shigeki Matsumoto}$^{(b)}$, \\
{\bf Takeo Moroi}$^{(c, b)}$
and
{\bf Norimi Yokozaki}$^{(b)}$

\vspace{1.5cm}
{\it
$^{(a)}${ICRR, University of Tokyo, Kashiwa, Chiba 277-8568, Japan} \\
$^{(b)}${Kavli IPMU, University of Tokyo, Kashiwa, Chiba 277-8568, Japan} \\
$^{(c)}${Department of Physics, University of Tokyo, Tokyo 113-0033, Japan} \\
}

\vspace{1.5cm}
\abstract{
The AMS-02 collaboration has recently reported an excess of cosmic-ray positron fractions, which is consistent with previous results at PAMELA and Fermi-LAT experiments. The result indicates the existence of new physics phenomena to provide the origin of the energetic cosmic-ray positron. We pursue the possibility that the enhancement of the positron fraction is due to the decay of gravitino dark matter. We discuss that such a scenario viably fits into the models in which the soft SUSY breaking parameters are dominantly from gauge-mediation mechanism with superparticle masses of around 10~TeV. Our scenario is compatible with 126~GeV Higgs boson, negative searches for SUSY particles, and non-observation of anomalous FCNC processes. We also point out that the scenario will be tested in near future by measuring the electric dipole moment of the electron and the lepton flavor violating decay of the muon.}

\end{center}
\end{titlepage}
\setcounter{footnote}{0}

\section{Introduction}

The AMS-02 collaboration has recently released their first result of the cosmic-ray positron fraction~\cite{AMS02}. The anomalous excess over the expectation based on the simple comic-ray propagation models has been seen again, and its energy spectrum is consistent with the previous results of the PAMELA~\cite{Adriani:2008zr} and Fermi-LAT~\cite{FermiLAT:2011ab} experiments. The AMS-02 collaboration has also reported a fact that the positron flux at the energy region of the excess shows no appreciable anisotropy to date. This fact may indicate that the positrons are not due to some astrophysical activities on the galactic plane but are from exciting dark matter phenomena in the halo, though it is clearly premature to make a definite statement because of limited statistics.

When the excess of the positron fraction is interpreted as a dark matter signal, a decaying gravitino with its mass being ${\cal O}(1)$\,TeV could be a promising candidate for dark matter~\cite{Ibarra:2007wg,Ishiwata:2008cu} (for earlier discussions on decaying gravitino dark matter see references \cite{Buchmuller:2007ui}). The lifetime of the gravitino is then required to be ${\cal O}(10^{26})$ sec., which is realized by introducing a tiny $R$-parity violation. The introduction of this violation is in fact favored from the viewpoint of cosmology; we can evade the serious gravitino problem~\cite{BBN} because the required violation is large enough to allow other sparticles to decay before the era of Big-Bang Nucleosynthesis (BBN). On the other hand, the violation is small enough not to wash out the baryon asymmetry of the universe created in the early universe, so that models with the decaying gravitino are consistent with the successful leptogenesis scenario~\cite{leptogenesis}.

In this letter, we revisit the model of decaying gravitino dark matter in light of the recent results of the AMS-02 experiment and the discovery of the Higgs particle at the LHC experiments~\cite{Aad:2012tfa, Chatrchyan:2012ufa}. As we show, the desirable gravitino mass and the observed Higgs boson mass at around $126$\,GeV can be explained simultaneously in the models of direct gauge mediation with the messenger masses just below  the scale of the Grand Unified Theory (GUT). We also show that the gravitino mass at around ${\cal O}(1)$\,TeV allows us to construct a simple model of the $R$-parity breaking which is appropriate for the favored gravitino lifetime, ${\cal O}(10^{26})$\,sec.

This letter is organized as follows. In section\,\ref{sec:AMS02}, we reassure that the decaying gravitino with a mass at around 1\,TeV well fits the anomalous excess of the positron fraction observed at the AMS-02. In section\,\ref{sec:model}, we discuss the models with gauge mediation at the GUT scale which explain both the favored gravitino mass of ${\cal O}(1)$\,TeV and the observed Higgs boson mass, $m_h \simeq 126$\,GeV. We also discuss the flavor changing neutral current and $CP$-violation in the lepton sector. In section\,\ref{sec:rpvmodel}, we  give a simple model of $R$-parity breaking which leads to the appropriate gravitino lifetime. The final section is devoted to our conclusions.

\section{Positron Fraction}
\label{sec:AMS02}

We first discuss the excess of positron fractions reported by the AMS-02 collaboration and its implication to the decaying gravitino dark matter. There are several ways to introduce R-parity violating interactions which make the gravitino meta-stable. In this letter, we consider the simplest possibility, namely the violation through the $LH_u$ operator as an example, where $L$ and $H_u$ are the lepton doublet and the up-type higgs superfields, respectively. The relevant part of the superpotential and soft SUSY breaking terms are therefore given by as follows;
\begin{eqnarray}
W &=&
\mu H_u H_d + \mu'_i H_u L_i,
\\
\mathcal{L}_{\rm soft} &=&
(B \mu \, H_u H_d + B'_i \mu'_i H_u \tilde{L}_i + h.c.)
- \tilde{L}^\dag_i m_L^2{}_{ij} \tilde{L}_j - m_{H_d}^2 |H_d|^2,
\nonumber
\end{eqnarray}
where $H_d$ is the down-type higgs superfield.
Then, the gravitino dark matter decays into a $Z$ boson plus a neutrino, a Higgs boson plus a neutrino, and a $W$ boson plus a charged lepton with the relative ratio of about 1:1:2, as was explicitly shown in reference~\cite{Ishiwata:2008cu}.

We next summarize our procedure to calculate the positron fraction $R=\Phi_{e^+}/(\Phi_{e^+} + \Phi_{e^-})$, where $\Phi_{e^+}$ and $\Phi_{e^-}$ are positron and electron fluxes, respectively.
The fluxes consist of the contribution from the decaying gravitino dark matter and the background contribution: $\Phi_{e^+}=[\Phi_{e^+}]_{\rm DM}+[\Phi_{e^+}]_{\rm bkg}$.
For the contribution from the dark matter, we have solved the diffusion equation in order to take account of the effect of electron/positron propagations inside our galaxy.
The energy spectrum of the electron/positron from the dark matter $f_{e^\pm}$ evolves as~\cite{Baltz:1998xv}
\begin{eqnarray}
\frac{\partial f_{e^\pm}(E, \vec{r})}{\partial t}
= K(E) \, \left[ \nabla^2 f_{e^\pm}(E, \vec{r}) \right]
+ \frac{\partial}{\partial E} \left[ b(E) \, f_{e^\pm}(E, \vec{r}) \right]
+ Q(E, \vec{r}),
\label{eq: diffusion equation}
\end{eqnarray}
where the function $K$ is expressed as $K = K_0 \, E_{\rm GeV}^\delta$ with $E_{\rm GeV}$ being the energy in units of GeV and $b = 1.0 \times 10^{-16} \, E_{\rm GeV}^2$ [GeV/sec].
In our numerical calculation, we have fixed the parameters $\delta$ and $K_0$ to be 0.55 and $5.95 \times 10^{-3}$ [kpc$^2$/Myr], respectively. The diffusion zone is assumed to be a cylinder with a radius of 20 [kpc] and a half-height of 1 [kpc], and we set $f_{e^\pm} = 0$ at the boundary.
The fluxes from the dark matter are then obtained as $[\Phi_{e^\pm}]_{\rm DM}(E)=(c/4\pi)f_{e^\pm}(E,\vec{r}_{\odot})$.

The source term $Q$ expressing the production of primary electrons/positrons from decaying gravitino dark matters is given by the following formula,
\begin{eqnarray}
Q(E, \vec{r}) =
\frac{\rho_{\rm DM}(\vec{r})}{m_{3/2} \, \tau_{3/2}}
\left[ \frac{dN_{e^\pm}}{dE} \right]_{\rm decay},
\end{eqnarray}
where $\tau_{3/2}$ and $m_{3/2}$ are the lifetime and the mass of the gravitino, respectively. In addition, $\rho_{\rm DM}$ is the dark matter mass density of our galaxy for which we adopt the NFW profile $\rho_{\rm DM}(\vec{r}) = \rho_\odot (r_\odot/r) (r_c + r_\odot)^2/(r_c + r)^2$~\cite{NFW}, where $\rho_\odot\simeq 0.4$ [GeV/cm$^3$] is the local halo density, $r_c \simeq 20$ [kpc] is the core radius, and $r_\odot \simeq 8.5$ [kpc] is the distance between the galactic center and our solar system. In the above expressions, $[dN_{e^\pm}/dE]_{\rm decay}$ is the energy distributions of the electron/positron from single decay process, and are calculated by using PYTHIA6 package~\cite{Pythia6.4}.

The background contributions are approximated to the form $[\Phi_{e^\pm}]_{\rm bkg}=C_{\pm}\cdot(E_{\rm GeV})^{\delta_\pm}$, where $[\Phi_{e^\pm}]_{\rm bkg}$ is in the unit of [GeV cm$^2$ sec sr]$^{-1}$.
Parameters for electrons are determined with fitting against the result of total electron flux reported by PAMELA collaboration~\cite{Adriani:2011xv} to be $(C_-,\delta_-)=(0.035,-3.24)$.
For positrons, $C_+=0.088C_-$ and $\delta_+={\delta_-}-0.26$ are used.

\begin{figure}[t]
\begin{center}
\includegraphics[width=0.8\hsize]{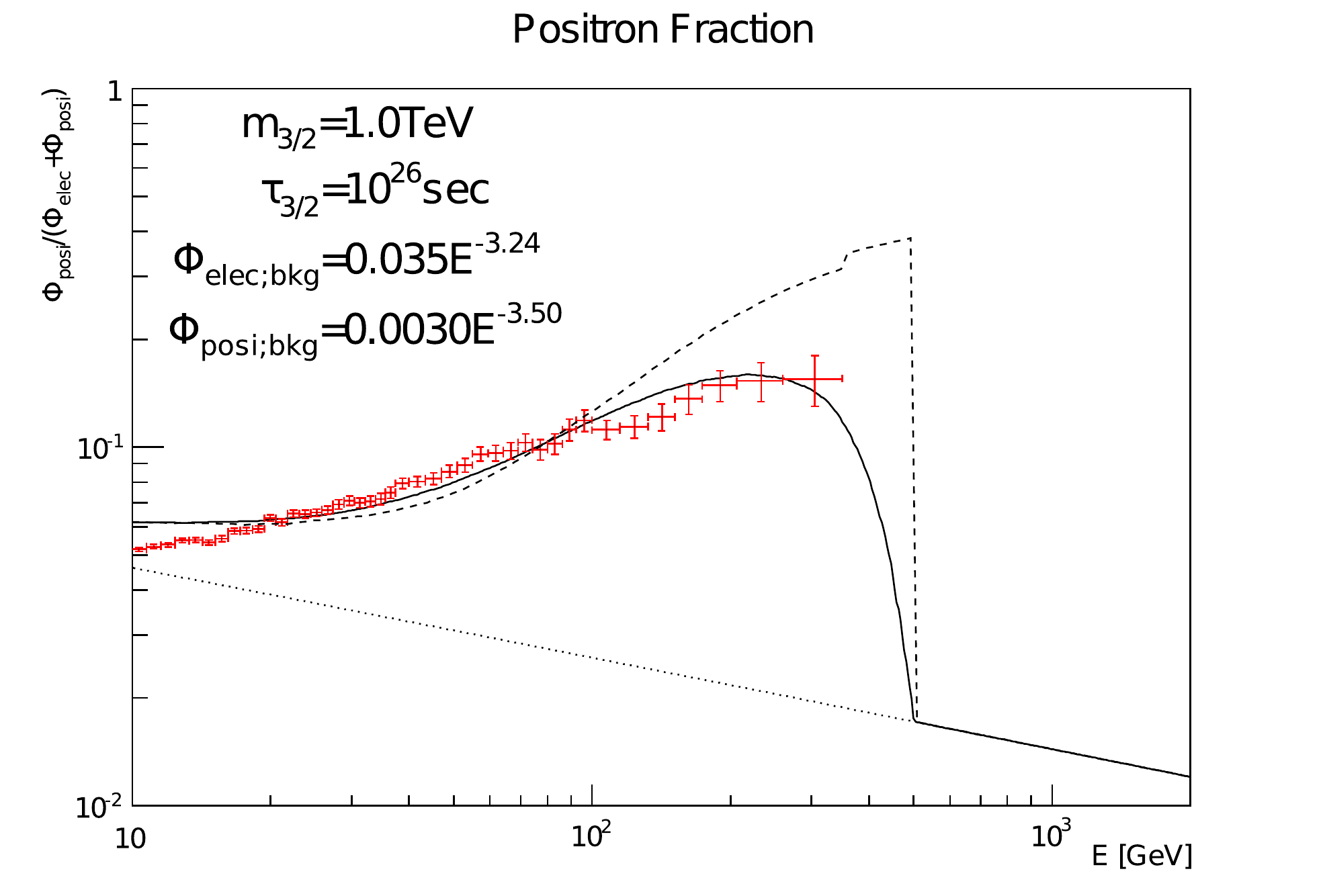}
\caption{\sl \small The positron fraction expected in the decaying gravitino dark matter scenario with the bilinear term of the R-parity violation.
The solid line corresponds to the $L_2H_u$ interaction, while the dashed line to the $L_1H_u$.
As a reference, the fraction without dark matter contribution is drawn with a dotted line.
The recent results from the AMS-02 experiment is shown with the red data points.
}
\label{fig:fraction}
\end{center}
\end{figure}

The result is shown in Fig.~\ref{fig:fraction}. 
As a benchmark point, the mass and lifetime of the dark matter are fixed as $(m_{3/2},\tau_{3/2})=(1.0{\,\rm TeV}, 10^{26}{\,\rm sec})$.
Two patterns of the gravitino decay are shown: the solid line (the dashed line) shows the case where the gravitino exclusively decays into the second (first) generation leptons, i.e., R-parity is broken only in the second (first) generation leptons.

First it should be noted that the data points below 10 GeV are not taken into consideration because the fraction in the region is subjected to the solar modulation~\cite{Maccione:2012cu}.
Also, fractions around 10 GeV could be modified by the change of the background contribution.
Considering these two uncertainties, the AMS-02 results are explained well by the solid line, i.e., when the gravitino decays into the second generation.
On the contrary, when the decay is into the first generation leptons, the slope becomes steeper due to the electrons produced primarily, and it is difficult to explain the AMS-02 result.

Decays of the gravitino through the $LH_u$ operator, on the other hand, also produce other particles such as anti-protons and gamma-rays.
Furthermore, high-energy electrons from the decay of the gravitino produce gamma-rays through the inverse Compton scattering with cosmic microwave backgrounds and infra-photons from star lights~\cite{Ishiwata:2009dk}.
Since no excesses are reported so far at observations of these particles, models involving the decaying gravitino are severely constrained.
First, let us consider the constraint from the observation of cosmic-ray anti-protons.
According to reference~\cite{Garny:2012vt}, with use of the most conservative limit (dark matter decay only), it turns out that the lifetime of the gravitino should be longer than 10$^{26}$ sec. when the gravitino mass is around 1~TeV.
If we used shallower profiles than the NFW one, the limit is expected to be milder~\cite{Evoli:2011id}. Next, we consider the constraint from gamma-ray observations. According to the most conservative limit (dark matter decay only) in reference~\cite{Cirelli:2012ut}, the lifetime has to be longer than about 10$^{26}$ sec. In any case, the lower limit on the lifetime is almost the same as that required to explain the excess of positron fractions. An attractive solution to this problem is the use of the $LL\bar E$ operator instead of $LH_u$, where we do not have  to worry about the constraint from the observation of cosmic-ray anti-protons, while that from gamma-ray observations becomes much milder than the case of $LH_u$~\cite{Shirai:2009fq}.

Finally, let us comment on the mass of the gravitino.
The decaying gravitino should be heavier than 520\,GeV in order to explain the rightmost data point of the AMS-02 result, which is for the energy of 260--350\,GeV.
Meanwhile, if the gravitino mass is heavier than 1\,TeV, the lifetime which can explain the result tends to be shorter than $10^{26}$ sec.\ since the number density of the gravitino is smaller.
This is somewhat disfavored for the constraints from the gamma-ray and anti-proton fluxes.
The AMS-02 result indicates that the mass of the decaying gravitino dark matter should be $\sim1$\,TeV, which provides precious information to model building behind the physics of the dark matter.

%
%
%
%
\section{Gravitino Dark Matter in Gauge Mediation}
\label{sec:model}

In the previous section, we have seen that the AMS-02 result is well
explained in the model with unstable gravitino dark matter with the
mass of $\sim 1\ {\rm TeV}$.  However, such a model is usually
severely constrained by the flavor violating processes, like the
$K$-$\bar{K}$ mixing, lepton flavor violating (LFV) processes, and so
on.  In particular, if the gravity mediation contribution dominates
the soft SUSY breaking sfermion masses, the off-diagonal elements of
the sfermion mass-squared matrices are expected to be unsuppressed
compared to the diagonal ones.  If so, our scenario is not compatible
with the constraints from the flavor violating processes.  In addition, in the light of the
recently observed Higgs mass, the scenario gives too small Higgs mass
(unless the tri-linear scalar coupling is relatively large); the stop masses should be as large as $\sim 10\ {\rm TeV}$ to
realize the lightest Higgs mass as large as 126 GeV \cite{Giudice:2011cg}.

If the gauge-mediation contribution gives an extra contribution to
the soft SUSY breaking parameters, the above problems may be solved.
In particular, if the gauge mediation contributions to the sfermion
masses are significantly larger than the gravitino mass and are as large
as $\sim 10$ TeV, we obtain viable scenario as we see below.
 
We parameterize the messenger sector as
\begin{eqnarray}
W_{\rm mess} = (M_{\rm mess} + F_{\rm mess} \theta^2) \Psi_i \bar{\Psi}_i,
\end{eqnarray}
where $\Psi_i$($\bar{\Psi}_i$) are messenger fields in fundamental
(anti-fundamental) representation of $SU(5)$ GUT gauge group. The
indices $i$ runs to $1 -N_{\rm mess}$, with $N_{\rm mess}$ being the
number of the messengers. The gravitino mass is bounded from below:
\begin{eqnarray}
  m_{3/2} \geq \frac{F_{\rm mess}}{\sqrt{3} M_P}.
\end{eqnarray}
The equality is satisfied in the direct gauge mediation models; in
such a model, the messenger scale is around the GUT scale taking
$m_{\rm soft} \sim 10$ TeV.

The bino, wino and gluino masses are given by
\begin{eqnarray}
  M_{\tilde{B}} &\simeq& 
  N_{\rm mess} \frac{\alpha_1}{4\pi} 
  \left( \frac{F_{\rm mess}}{M_{\rm mess}} \right), 
  \\
  M_{\tilde{W}} &\simeq& 
  N_{\rm mess} \frac{\alpha_2}{4\pi} 
  \left( \frac{F_{\rm mess}}{M_{\rm mess}} \right),  
  \\
  M_{\tilde{g}} &\simeq& N_{\rm mess} 
  \frac{\alpha_3}{4\pi} 
  \left( \frac{F_{\rm mess}}{M_{\rm mess}} \right),
\end{eqnarray}
where $\alpha_1$, $\alpha_2$ and $\alpha_3$ are gauge coupling
constant squareds of $SU(3)_C$, $SU(2)_L$ and $U(1)_Y$ divided by
$4\pi$, respectively.  (Here, we take a normalization of $U(1)_Y$ as
the $SU(5)$ GUT normalization.)  In addition, the messenger-scale
values of the sfermion masses are
\begin{eqnarray}
  m_{Q}^2 &\simeq& 
  \left( \frac{4}{3} \alpha_3^2 + \frac{3}{4} \alpha_2^2
    + \frac{1}{60} \alpha_1^2  
  \right) 
  \frac{N_{\rm mess}}{8\pi^2} \left( \frac{F_{\rm mess}}{M_{\rm mess}} \right)^2, 
  \\
  m_{\bar{U}}^2 &\simeq& 
  \left(\frac{4}{3} \alpha_3^2+ \frac{4}{15} \alpha_1^2 \right) 
  \frac{N_{\rm mess}}{8\pi^2}   
  \left( \frac{F_{\rm mess}}{M_{\rm mess}} \right)^2,
  \\
  m_{\bar{D}}^2 &\simeq& 
  \left(\frac{4}{3} \alpha_3^2+ \frac{1}{15} \alpha_1^2  \right)
  \frac{N_{\rm mess}}{8\pi^2} 
  \left( \frac{F_{\rm mess}}{M_{\rm mess}} \right)^2,
  \\
  m_{L}^2 &\simeq& 
  \left( \frac{3}{4} \alpha_2^2+ \frac{3}{20} \alpha_1^2 \right)
  \frac{N_{\rm mess}}{8\pi^2}  
  \left( \frac{F_{\rm mess}}{M_{\rm mess}} \right)^2,
  \\
  m_{\bar{E}}^2 &\simeq& 
  \frac{3}{5} \alpha_1^2 \frac{N_{\rm mess}}{8\pi^2}
  \left( \frac{F_{\rm mess}}{M_{\rm mess}} \right)^2,
  \label{eq:sfermion}
\end{eqnarray}
where $Q, \bar{U}, \bar{D}$ are squarks and $L, \bar{E}$ are sleptons.
The soft masses of the up- and down-type Higgses ($H_u$ and $H_d$) are
same as $m_{L}^2$.

The mass spectrum we anticipated can be easily obtained.  For
instance, taking $M_{\rm mess}=10^{15}$ GeV, $F_{\rm mess}/M_{\rm
  mess}=3000\ {\rm TeV}$, and $N_{\rm eff}=1$, the low-energy values
of gluino and squark masses are abut 20 TeV while the slepton masses
are $\sim 10$ TeV.  Furthermore, assuming the direct gauge mediation,
the gravitino mass becomes about 1 TeV.  (Notice that the
gravitino mass as large as 1 TeV can be realized with smaller value of
$M_{\rm mess}$ if we do not consider the direct gauge mediation.)

Let us estimate the rates of flavor and CP violations which are
important probes of the SUSY particles even if the masses of
superparticles are quite large \cite{Gabbiani:1996hi, Moroi:2013sfa,
  McKeen:2013dma}.  First, we consider the leptonic ones on which the
experimental bounds are expected to be drastically improved.  Even
though all the superparticles are relatively heavy, the rates of LFV processes may become sizable.  This is
because the off-diagonal elements of the sfermion mass-squared
matrices are expected to be of $O(m_{3/2}^2)$, which are $\sim 1\ \%$
of the diagonal elements in the present setup.  If the masses of
sleptons, Higgsinos, and the electroweak gauginos are $10\ {\rm TeV}$
while the 1-2 elements of the slepton mass-squared matrices are
$10^{-2}$ of the diagonal elements, for example, $Br(\mu\rightarrow
e\gamma)$ is estimated to be $1\times 10^{-15}$, $6\times 10^{-15}$,
and $2\times 10^{-14}$, for $\tan\beta=5$, $10$, and $20$,
respectively.  Thus the rate of the $\mu\rightarrow e\gamma$ process
can be below the current bound.  In addition, using the fact that
$Br(\mu\rightarrow e\gamma)$ is approximately proportional to
$\tan^2\beta$, $Br(\mu\rightarrow e\gamma)$ may be within the reach of
MEG-upgrade experiment (which is expected to reach $Br(\mu\rightarrow
e\gamma)\simeq 6\times 10^{-14}$ \cite{Baldini:2013ke}) in particular
when $\tan\beta$ is large.

Another important check point is the so-called $\epsilon_K$ parameter,
which often gives the most stringent constraint on supersymmetric
flavor and CP violations.  Assuming that off-diagonal elements of the
squark mass-squared matrices are given by $(1\ {\rm TeV})^2$, for
example, the SUSY contribution to the $\epsilon_K$ parameter becomes
smaller than the present bound ($1\times 10^{-3}$ \cite{Brod:2011ty,
  Beringer:1900zz}) if the mass scale of the colored superparticles is
larger than $\sim 20\ {\rm TeV}$. (Here, the phases in the MSSM
parameters are tuned to maximize the SUSY contribution to $\epsilon_K$.)
Such a value of the colored superparticle masses can be easily
realized as we have mentioned.

We also comment on the electric dipole moment (EDM) of the electron.
Assuming that the supergravity contribution provides the $B$ parameter
of $O(m_{3/2})$, we expect that the CP violating phase in the $\mu$
parameter (in the bases where the VEVs of the Higgs bosons and the
gaugino masses are real) is sizable.  Taking the masses of non-colored
superparticles to be $10\ {\rm TeV}$, for example, the electron EDM is
estimated to be $0.3\times 10^{-27}$, $0.6\times 10^{-27}$, and
$1\times 10^{-27}\ e{\rm cm}$ for $\tan\beta=5$, $10$, and $20$,
respectively, which are around the current bound
\cite{Beringer:1900zz}.  (Here, we have tuned the phase of the $\mu$
parameter to maximize the electron EDM.)  The future experiments,
which may improve the bound by $\sim 3$ orders of magnitude
\cite{Vutha:2009ux}, have a good chance to observe the electron EDM.

Now, we estimate the size of $\mu'$ and soft masses required for the
gravitino decay with $\sim 10^{26}$ sec. In the case that R-parity
is broken by bi-linear operator, we need a small VEV of a left-handed
sneutrino, inducing chargino, neutralino/lepton mixing. Then, the
gravitino decays into the SM gauge bosons and leptons through the
mixing. The sneutrino get a VEV trough the potential
\begin{eqnarray}
\mathcal{L}_{\rm soft} &\ni& \tilde{B}_i^2 \tilde{L}_i H_u - m_{i}^2 \tilde{L}_i H_d^* + h.c. - m_{\rm GMSB}^2 |\tilde{L}|^2, \label{eq:pot}
\end{eqnarray}
where $\tilde{B}_i^2 = -\epsilon_i B \mu + \epsilon_k B_k' \mu
\delta_{ik}$ and $m_i^2= (\delta m_{L}^2)_{ik} \epsilon_k^* -
\epsilon_i \delta m_{H_d}^2$.  The rotation parameters $\epsilon_i$
are determined by the ratio of $\mu$ and $\mu_i'$ as
$\epsilon_i=\mu_i/\mu$, and $m_{\rm GMSB}^2=m_{L}^2$ are the soft
masses induced by gauge mediation. Other mass parameters, $B$, $B'$,
$\sqrt{|\delta m_L^2|}$ and $\sqrt{|\delta m_{H_d}^2|}$ of $\sim 1\
{\rm TeV}$ are induced by the supergravity effects. The sneutrino VEV
is given by
\begin{eqnarray}
\kappa_i \equiv \frac{\left<\tilde \nu_i\right>}{v} \simeq  (\tilde{B}_i^2 \sin\beta - m_{i}^2 \cos\beta)/m_{\rm GMSB}^2.
\end{eqnarray}
For the gravitino of $\sim 1$ TeV, the life-time of $10^{26}$ sec.\
is explained by $\kappa_i \sim 10^{-10}$~\cite{Ishiwata:2008cu}. This
requires $\epsilon_i$ to be $\sim 10^{-9}$, i.e., the coefficient of
the R-parity violating bi-linear term $\mu'_i$ should be $\sim
10^{-5}$ GeV ($\mu \sim 10$ TeV in the parameter space of our
interest). This $\mu'_i \sim 10^{-5}$ GeV can be explained
consistently with the seesaw mechanism; $\mu'$ is induced by the small
vacuum expectation value of the right-handed sneutrino through the
operator $L H_u \left< \bar{N}_R \right>$.  An explicit model
realizing the small $\left< \bar{N}_R \right>$ is discussed
below. Note that the NLSP also decays into SM particles through the
mixing with the life-time much less than 1 sec, and hence, the BBN
constraint can be avoided~\cite{BBN}.

%
%
%
\section{A Model of R-parity Violation}
\label{sec:rpvmodel}

In the previous section, we have shown that the small value of the
R-parity violating parameter of $\mu' = O(10^{-5})$\,GeV is needed to
explain the AMS-02 result.  Here, we show that such a value of the
$\mu'$ parameter may arise in the model with a discrete R-symmetry, as
we show below.

Let us consider a model with a discrete $R$-symmetry, ${\mathbb Z}_{5R}$
(see Table.\,\ref{tab:charge}).
With the given charge assignments, the relevant superpotential terms are given by,
\begin{eqnarray}
\label{eq:model}
 W &=& y_u {H_u}\, \mathbf{10}\,\mathbf{10}
 +  y_{d,e} {H_d}\mathbf {10}\,{\mathbf 5}^*
  + y_\nu {H_u}\, {\mathbf 5}^*\, \bar{N}_R+ \mu H_uH_d
  + \,y_{M} \phi  \bar{N}_R^2/2    \nonumber\\
&&
- \frac{ c_4}{M_{\rm PL}^2} \phi^4\bar{N}_R 
+ \frac{ c_7}{M_{\rm PL}^4} \phi^7
  + y_X X\left(\phi^2  + c_H\frac{\mu}{M_{\rm PL}}H_uH_d-c_W\frac{\vev{W_0}}{M_{\rm PL}}\right) 
    \ ,
\end{eqnarray}
where $\phi$ and $X$ are gauge singlets and $y$'s and $c$'s are dimensionless constants.%
\footnote{
Here, we absorbed the allowed mass term of $\phi^2$ by the shift of $X$.
We also assume that the size of $\mu$ is controlled by another symmetry than ${\mathbb Z}_{5R}$
symmetry.
}
We assume here that there is a spontaneous discrete $R$-symmetry breaking sector which generates
$m_{3/2} = \vev{W_0}/M_{\rm PL}^2 \neq 0$.%
\footnote{
For example, the discrete $R$-symmetry is spontaneously broken 
in the supersymmetric Yang-Mills theories by the gaugino condensations\,\cite{gauginocondensation}.
If $X$ appears in the gauge kinetic functions of the Yang-Mills theories,
we need several gaugino condensations to stabilize $X$.
}
It should be noted that the above superpotential breaks $U(1)_{B-L}$ and $R$-parity explicitly, 
although the usual $R$-parity violating terms $LL\bar E$, $LQ\bar{D}$, $\bar{U}\bar{U}\bar{D}$ 
and $LH_u$ are suppressed by ${\mathbb Z}_{5R}$.

From the above superpotential, we find a stable vacuum at around,
\begin{eqnarray}
\label{eq:vacuum}
\vev{\phi} &=& \left(c_W\frac{\vev{W}}{M_{\rm PL}}\right)^{1/2} = (c_Wm_{3/2}M_{\rm PL})^{1/2}\ ,\\
\vev{\bar{N}_R} &=& \frac{c_4}{y_M}
\frac{\vev{\phi}^3}{M_{\rm PL}^2}  = 
\frac{c_4}{y_M}
\left(\frac{c_W^3m_{3/2}^3}{M_{\rm PL}}\right)^{1/2}\ ,\nonumber\\
\vev{X} & = & - \frac{y_M\vev{\bar{N}_R}^2}{4y_X\vev{\phi}}
+ \frac{2c_4}{y_XM_{\rm PL}^2} \vev{\phi}^2\vev{\bar{N}_R}
+ \frac{7c_7}{2y_XM_{\rm PL}^4} \vev{\phi}^5 \sim m_{3/2} \left(\frac{m_{3/2}}{M_{\rm PL}}\right)^{5/2}\  .
\end{eqnarray}
For $c_W = {\cal O}(1)$, $c_4/y_M = {\cal O}(1)$ $m_{3/2} = {\cal O}(1)$\,TeV,  we find 
\begin{eqnarray}
\vev{\phi} &= &{\cal O}(10^{11})\,{\rm GeV}\ , \nonumber\\
\vev{\bar{N}_R} &=&  {\cal O}(10^{-5}){\rm GeV}\ .
\end{eqnarray}
At around this vacuum, we obtain the masses of the right-handed neutrino,
\begin{eqnarray}
 M_N \propto \vev{\phi} = {\cal O}(10^{11})\,{\rm GeV}\ , 
\end{eqnarray}
which are appropriate for the see-saw mechanism\,\cite{seesaw}.
Furthermore, we also obtain the effective bi-linear R-parity violating term,
\begin{eqnarray}
 \mu' = \vev{\bar{N}_R} &=&  {\cal O}(10^{-5})\,{\rm GeV}\ ,
\end{eqnarray}
which is appropriate for the decaying gravitino dark matter discussed in the previous section.

In addition to the superpotential terms in Eq.\,(\ref{eq:model}), 
there are $R$-parity breaking terms such as,
\begin{eqnarray}
 W = \frac{1}{M_{\rm PL}^3}\phi \vev{W_0} H_u \, {\mathbf 5}^*\ ,
\end{eqnarray}
which gives a similar contribution to the  $LH_u$ term from the $\vev{\bar{N}_R} L H_u$ term.
The trilinear $R$-parity breaking terms have, on the other hand, the charge $-1$ under $\mathbb{Z}_{5R}$,
and hence, they are proportional to $(m_{3/2}/M_{\rm PL})^{3/2} = {\cal O}(10^{-22})$.
Thus, the trilinear $R$-parity breaking terms generated by the vacuum expectation values
in Eq.\,(\ref{eq:vacuum}) are negligibly small.

\begin{table}[t]
\caption{\sl\small
The charge assignments under the discrete ${\mathbb Z}_{5R}$ symmetry are presented here.
We have used $SU(5)$ GUT representations for the MSSM matter fields,
i.e. $\mathbf{10} = (Q_L,\bar{U}_R,\bar{E}_R) $, ${\mathbf 5}^* = (\bar{D}_R,L_L)$
and $\bar{N}_R$ the right-handed neutrino.
}
\begin{center}
\begin{tabular}{|c||c|c|c|c|c|c|c|}
\hline
& $H_u$ & $H_d$ & $\mathbf{10}$ & ${\mathbf 5}^*$& $\bar{N}_R $
& ${\phi} $ & X 
\\
\hline
${\mathbb Z}_{5R}$&$1$&  $1$ & $3$ & $3$ & $3$ & $1$&  $0$
\\
\hline
\end{tabular}
\end{center}
\label{tab:charge}
\end{table}%

\section{Summary}

In this letter, we have shown that the anomalous increase of the
positron fraction recently observed by the AMS-02 experiment can be
well explained in the unstable gravitino scenario if $m_{3/2}\gtrsim
520\ {\rm GeV}$ and $\tau_{3/2}\simeq 10^{26}$ sec.  Even with such a value
of the gravitino mass, we argued that the dangerous SUSY contribution
to the flavor violating processes may be suppressed enough if the
gauge mediation mechanism provides the dominant contribution to the
sfermion masses of $\sim 10\ {\rm TeV}$.  We have also proposed a
model to give the R-parity violating interaction required to explain
the AMS-02 signal.

Here, we concentrated on the case where the gravitino decays into
lepton and electroweak gauge boson (or Higgs) with the interaction
given in Eq.\ \eqref{eq:pot}.  With other types of
R-parity violating interaction, however, gravitino may decay into
different final states.  In particular, if the $LL\bar E$-type R-parity
violating operator exists in the superpotential, the gravitino decays
into purely leptonic final state, with which the constraints from the
anti-proton flux can be easily avoided.  More detailed analysis on the related issues will
be given elsewhere.

\vspace{1.0cm}

\noindent
{\bf Acknowledgments}
\vspace{0.1cm}\\
\noindent
The authors thank the Yukawa Institute for Theoretical Physics at Kyoto University, where this work was initiated during the YITP workshop on "LHC vs. Beyond the Standard Model (YITP-W-12-21)," March 19--25, 2013, and also acknowledge the participants of the workshop for very active discussions. 
This work is supported by the Grant-in-Aid for Scientific research from the Ministry of Education, Science, Sports, and Culture (MEXT), Japan (Nos. 22244021, 23104008, 23740169, 24740151, and 60322997), and also by the World Premier International Research Center Initiative (WPI Initiative), MEXT, Japan. The works of S.I. and N.Y. are supported by JSPS Research Fellowships for Young Scientists.


\end{document}